\documentclass[aps, pra, showpacs, preprint]{revtex4-2}
\usepackage{lineno}

\usepackage{cmap}
\usepackage[utf8]{inputenc}
\usepackage[english]{babel}
\usepackage{amssymb,amsfonts,bm,mathtext,longtable,array,amsmath}
\usepackage{xcolor}
\usepackage{svg}

\usepackage{graphicx}
\graphicspath{{1/}}

\begin{document}

\title{Calibration features of a polarimetric backscattering setup with a beamsplitter} 

\author{V. Stefanov, B.P. Singh and A. Stefanov}
\affiliation{Institute of Applied Physics, University of Bern, Sidlerstrasse 5, CH-3012 Bern, Switzerland}

\begin{abstract}
\noindent We address the problem of calibrating a polarimetric backscattering setup that includes a beamsplitter, taking its optical properties into account. Using Fisher information, we justify a specific calibration configuration involving reference optical elements for the maximum likelihood method and demonstrate its effectiveness through Monte Carlo simulations. This approach leads to higher precision, as the described set of measurements allows the beamsplitter to be treated as an integral part of the setup, without requiring separate processing or reconfiguration of the setup for the beamsplitter calibration. Furthermore, we emphasize the significance of considering the beamsplitter's optical properties by proposing a test to estimate the error that arises from neglecting these properties during the calibration procedure.
\end{abstract}

\maketitle

\section{Introduction}

Polarimetry is a powerful experimental tool for characterizing the optical properties of objects of interest. It has numerous applications in fields such as biomedical research \cite{ghosh2011tissue,ulaby1990radar}, astronomy \cite{serkowski1974polarimeters,hough2007new}, and geology \cite{egan2003optical}, among others \cite{schmullius1997review}. The distinctly different polarization features of an object's components provide excellent contrast and high sensitivity. Today, polarimetry-based devices are among the most promising tools for recognizing various types of cancer in vivo or during surgical procedures \cite{8744376,louie2021constructing,qi2017mueller}.

The full polarimetric properties of samples can be probed either in transmission or reflection. The typical reflection configuration is only sensitive to the surface of the sample \cite{vitkin2015tissue,li2022polarimetric,guo2013study}. In this work, we focus on the so-called backscattering configuration \cite{Jain:20,jain2021backscattering,arteaga2014elementary}, which allows for the study of the internal scattering properties of a sample, rather than just its surface. This makes it particularly useful for applications beyond histological analysis \cite{jain2021backscattering}. A set of points with equal polarimetric properties can delineate boundaries between regions with different internal structures, such as the boundary between anisotropic brain tissue and tumors \cite{rodriguez2022polarimetric,jain2021backscattering}. When combined with time-resolved detection, this method potentially enables the study of tissue samples layer by layer to the full penetration depth \cite{Stefanov:23}. The backscattering setup uses focused illumination, implying a short working distance. As a result, the illumination light is typically separated from the collected backscattered light using a beamsplitter (BS).

The technical essence of polarimetry lies in defining the so-called Mueller matrix for each point of the studied object. The $4 \times 4$ Mueller matrix provides complete information about the polarization properties, enabling the prediction of all possible polarizations of scattered light at a given point.

The definition of a Mueller matrix requires conducting at least 16 measurements for each pair of the four possible input and output polarizations (which are generated by a polarization state generator (PSG) and a polarization state analyzer (PSA), respectively). Our knowledge of the values of PSG and PSA is one of the key factors determining the accuracy of the setup. Several calibration methods are available to measure these values, including eigenvalue calibration methods (ECM) \cite{compain1999general,de2004general,macias2012eigenvalue}, the maximum likelihood calibration method (MLCM) \cite{hu2013maximum}, and others \cite{Chen:18,Meng:21}. However, these methods are generally not suitable for a backscattering polarimetric setup involving a BS, due to the need to account for its specific optical properties. Some variants of the ECM method \cite{JEOS:RP12004,lara2006axially} can be applied under certain assumptions about the Mueller matrix of the BS. Nevertheless, the BS is typically calibrated separately from the rest of the setup \cite{zhang2018characterization}, which still introduces errors in the final configuration of the system. In this paper, we present a configuration of reference elements that allows the calibration of the setup directly via MLCM, without the need for additional manipulation of the BS.

\begin{figure}[h]
\begin{minipage}[b]{0.5\linewidth}
  \centering
  \includegraphics[width=0.8\linewidth]{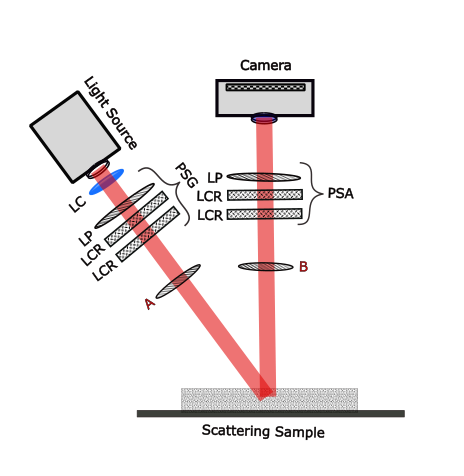}   
\end{minipage}%
\begin{minipage}[b]{0.5\linewidth}
  \centering
  \includegraphics[width=1\linewidth]{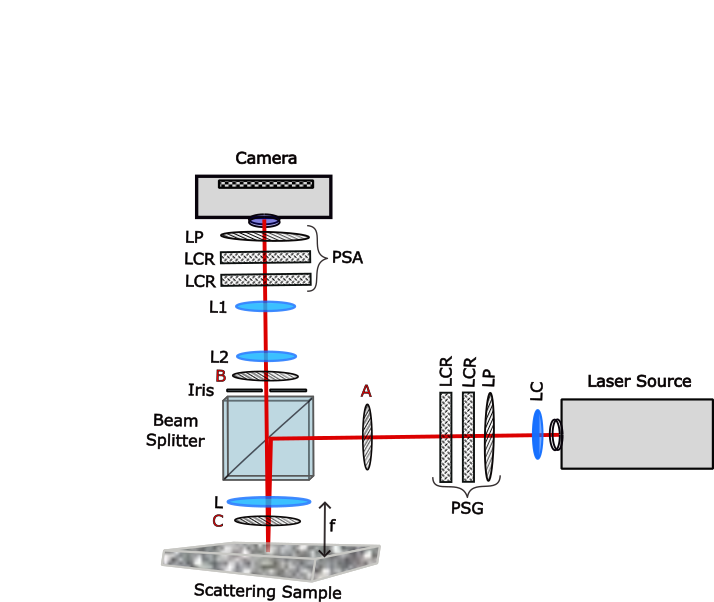} 
\end{minipage}
\newline
\begin{minipage}[b]{0.5\linewidth}
  \centering
  (a)
\end{minipage}%
\begin{minipage}[b]{0.5\linewidth}
  \centering
  (b)
\end{minipage}%
\caption{Two different principal schemes of a polarimetric back-scattering setup (a) reflective and (b) scanning.}
\label{ris:schemes}
\end{figure}

The paper is organized as follows: first, we provide a brief review of common calibration methods. Next, we describe a suitable calibration scheme for a BS setup and explain the features of using various optical elements. We then propose a test to estimate the errors of different calibration schemes, and, finally, we demonstrate the superiority of the suggested scheme through experimental validation, followed by a discussion.

\section{Common calibration methods}

We can collect the Stokes vectors generated by the PSG and PSA in matrices, denoted $W$ and $A$, respectively. This allows us to express the measured intensity values in a matrix $I$ and match the expression $I = AMW$ in the configuration without a BS (Type I in Fig.~\ref{ris:schemes}(a)) and $I = ATMBW$ in the configuration with a BS (Type II in Fig.~\ref{ris:schemes}(b)), where $M$, $B$, and $T$ are the Mueller matrices of the sample, the reflected side of the BS, and the transmitted side of the BS, respectively. Calibration involves retrieving the pair of matrices $A$ and $W$ for Type I, or $AT$ and $BW$ for Type II, from a set of measurements. For this purpose, reference optical elements with known or easily reconstructed Mueller matrices are used, such as linear polarizers ($P$), wave plates, mirrors, or other reflected samples ($R$).

Below, we briefly explain the essence of the eigenvalue calibration method (ECM) \cite{compain1999general} and the maximum likelihood calibration method (MLCM) \cite{hu2013maximum}.

\subsection{Eigenvalues Calibration Method}

According to the ECM, the four measured intensity matrices - two with reflected samples ($A R_1 W$, $A R_2 W$), one with a linear polarizer placed after the PSG ($A R_1 P W$), and one with a linear polarizer placed before the PSA ($A P R_1 W$) - are sufficient for calibration. The method is based on the fact that the eigenvalues of $W^{-1}R_1 W$ are the same as those of $R_1$. For this reason, calculating the eigenvalues of the expression $(A R_1 W)^{-1} (A R_1 P W)$ allows us to restore the Mueller matrix of the linear polarizer $P$ independently of its orientation angle $\theta$, while the expression $(A R_1 W)^{-1} (A R_2 W)$ allows us to restore the intermediate matrix $R_1^{-1} R_2$. Due to the presence of noise, we must apply an additional transformation, so we define the matrices $H_1(\theta)$ (as a function of $\theta$) and $H_2$:
\begin{eqnarray}
    H_1(\theta)&=&E \otimes P(\theta)-\left[(A R_1 W)^{-1} (A R_1 P W)\right]^T \otimes E, \label{H1}\\ 
    H_2&=&E \otimes R_1^{-1}R_2-\left[(A R_1 W)^{-1} (A R_2 W)\right]^T \otimes E,\label{H2}
\end{eqnarray}
where $\otimes$ means tensor product, $E$ is the identity matrix, $P(\theta)=J(\theta)P J(-\theta)$ is a Mueller matrix with the polarizer matrix $P=\tau_p\mathfrak{G}(1,0,0)$ and with a rotation matrix $J(\theta)$:
\begin{equation}\label{TB}
 \mathfrak{G}(x,y,z)=\frac{1}{2}\left[\begin{array}{cccc}
    1 & x & 0 & 0  \\
     x & 1 & 0 & 0  \\
     0 & 0 & y & z   \\
     0 & 0 & -z & y  
\end{array}\right],  \quad  J(\theta)=\left[\begin{array}{cccc}
    1 & 0 & 0 & 0  \\
     0 & \cos 2\theta & -\sin 2\theta & 0  \\
     0 & \sin 2\theta & \cos 2\theta & 0   \\
     0 & 0 & 0 & 1  
\end{array}\right].
\end{equation}
The Mueller matrices of reflected samples are parametrized as follows:
\begin{equation}\label{mm_of_reflected}
    R_{m}=R(\tau_{m},\psi_m,\Delta_{m}) \equiv  2\tau_m \mathfrak{G}(-\cos 2\psi_m,\sin 2\psi_m \cos \Delta_m,\sin 2\psi_m \sin \Delta_m),
\end{equation}
for $m=1,2$, where $\tau$ is a transmissivity, $\psi$ and $\Delta$ are ellipsometric angles. 

The linear mapping $K(\theta)$  is given by: 
\begin{equation}
    K(\theta)=H_1(\theta)^T H_1(\theta)+H_2^T H_2,
\end{equation}
and it allows for the definition of the orientation of the linear polarizer $\theta$ by minimizing the ratio of its smallest and second-smallest eigenvalues. The eigenvector of $K(\theta)$ corresponding to the smallest eigenvalue must be associated with the estimator of $W$, written in lexicographic order. Using the set $A P R_1 W$, $A R_1 W$, and $A R_2 W$ allows us to define $A$ in a similar manner.

\subsection{Maximum Likelihood Calibration Method}

Another possible way to calibrate a setup is through the maximum likelihood calibration method (MLCM). For a given set of measurements, the MLCM allows us to estimate the most likely state \cite{hradil20043}. To perform calibration using MLCM, we need to maximize the likelihood function $\mathfrak{L}(\zeta)$ over the set of all setup parameters $\{\zeta\}$: 
\begin{equation}
    \{\zeta\}=\arg\max_{\zeta} \mathfrak{L}(\zeta).
\end{equation}
Usually, it is more convenient to use the natural logarithm of the likelihood function, called the log-likelihood, $\mathfrak{l}(\zeta) = \log \mathfrak{L}(\zeta)$. The explicit form of the log-likelihood depends on the statistical distribution of the noise we are dealing with. In this paper, we assume the noise is Gaussian, and we present the expressions for the log-likelihood accordingly:
\begin{equation}
    \mathfrak{l}(\zeta)=-\frac{1}{2\sigma^2}\sum_s Tr\left[ (I_s-\beta_s A M_s W)\cdot (I_s-\beta_s A M_s W)^T\right], \label{log-likelihood}
\end{equation}
where $\sigma$ represents the standard deviation of the Gaussian noise, $s$ denotes the measurement number, $\beta_s$ stands for the effective transmittance coefficient, and $M_s$ is the Mueller matrix corresponding to the $s$-th measurement.

The assumption of Gaussian noise has limitations when dealing with pixel-dependent images at low intensities. Since each pixel follows Poisson statistics, deviations from a Gaussian distribution can occur at the scale of the photon count. Additionally, correlations between neighboring pixels must be considered, as they also influence the noise distribution at low intensities. These limitations may require the use of an alternative expression for the maximum likelihood function in place of Eq.~\eqref{log-likelihood}, although the MLCM would still remain applicable.

Both matrices, $A = \{\bm{S}_1, \dots, \bm{S}_4\}$ and $W^T = \{\bm{S}_5, \dots, \bm{S}_8\}$, contain four Stokes vectors in standard form, $\bm{S}_l = \{I_l, Q_l, U_l, V_l\}$. Since we cannot distinguish the transmittance input of each $s$-measurement part in the effective coefficient $\beta$ during numerical calculations, it is logical to normalize $A$ by $I_1$ and $W$ by $I_5$, incorporating them into $\beta$ explicitly.

Due to the linear dependence on $\beta_s$ in Eq.~\eqref{log-likelihood}, we can estimate them directly from the equation
\begin{equation}
    \frac{\partial{\mathfrak{l}}}{\partial{\beta_s}}=0.
\end{equation}
It is important to distinguish between the $\beta_s$ coefficients associated with different measurements, for example, when fitting data from several different orientations of a single polarizer. In this case, the expression for maximization can be rewritten as:
\begin{equation}\label{MLM}
    \mathfrak{l}(\zeta)=\sum_s\frac{\left(\sum\limits_{i_s} Tr\left[(A M_{i_s} W)\cdot I_{i_s}^T \right]\right)^2}{\sum\limits_{i_s} Tr\left[(A M_{i_s} W)\cdot (A M_{i_s} W)^T \right]},
\end{equation}
where we have $s$ sets of $i_s\geq1$ measurements with the equal staff, and the effective coefficient $\beta_s$ is equal to:
\begin{eqnarray}\label{coef_MLM}
    \beta_s=\frac{\sum\limits_{i_s} Tr\left[(A M_{i_s} W)\cdot I_{i_s}^T \right]}{\sum\limits_{i_s} Tr\left[(A M_{i_s} W)\cdot (A M_{i_s} W)^T \right]}.
\end{eqnarray}

\section{Calibration configuration for backscattering setup with a beamsplitter}

The ECM enables the reconstruction of only two generalized Mueller matrices, separated by polarizers at positions A and B in Fig.~\ref{ris:schemes}(b). This presents a challenge for setups involving a BS. Specifically, when calibrating both the PSA and PSG together with the BS via ECM, placing the polarizer between the sample and the BS (at position C in Fig.~\ref{ris:schemes}(b)) results in the light passing through the BS twice: first, immediately after the BS and then after reflection from the sample. The full Mueller matrix of the system can be written as $AP(-\theta)R_1 P(\theta)W$. This makes it impossible to determine the $H_i$ matrices in Eq.~\eqref{H1} and Eq.~\eqref{H2}, because the fundamental assumption for reconstructing $P(\theta)$ (and $R_1^{-1}R_2$) based on the equality of eigenvalues for $W^{-1}P(\theta)W$ and $P(\theta)$ does not hold here.

There are some modifications of the ECM tailored for BS configurations, but each relies on specific assumptions about the BS and a reference sample. For example, variants such as two- and three-step ECM \cite{JEOS:RP12004} require that the setup between the PSG and the PSA includes only retardance, with no depolarization, diattenuation, or polarizance. Another one, known as double-pass ECM \cite{lara2006axially}, assumes that the light reflected back through the polarization elements comes from a perfect mirror at normal incidence. Due to these constraints, ECM and its modifications can hardly be considered general calibration methods for polarimetric setups involving a BS.

However, the MLCM can handle any configuration of the system, including the case described above. To separate the PSG and PSA matrices from the sample, we need to place a reference optical element at position C in Fig.~\ref{ris:schemes}(b). For calibration with this auxiliary optical element $F$, set at angles $\theta_i$ and with sample $M$, the total expression for maximization via MLCM in this configuration takes the form in Eq.~\eqref{MLM}, with 
\begin{equation}\label{MIS}
M_{i_s} = F_s(-\theta_{i_s}) M F_s(\theta_{i_s}),
\end{equation}
where $F_s$ is oriented for $\theta_{i_s}$ as 
\begin{equation}\label{FS}
F_s(\theta_{i_s}) = J(\theta_{i_s}) F_s J(-\theta_{i_s}).
\end{equation}

An important feature of the MLCM is that it may not exhibit high sensitivity to variations in variables when their impact is small. This means that it can still produce reasonable calibration values, even in cases where the measurement set is poorly defined. In fact, the MLCM can find a solution even when some variables are nearly indistinguishable within the measurement set, due to the influence of initial conditions and noise in the intensity measurements.
To address this issue, it is essential to estimate the errors associated with the calibration parameters. One effective way to achieve this is by using the Fisher Information (FI) framework \cite{fisher1922mathematical}, which helps quantify the uncertainty in restoring variables as a function of experimental noise.

\subsection{Fisher information}

The FI is a powerful statistical tool that quantifies how much information an observable random variable carries about an unknown parameter of the probability distribution that models it. FI naturally arises in the context of the maximum likelihood estimator, which converges to the true parameter value through the normal distribution, when the number of observations tends to infinity \cite{jung2021optimal}. Then, the inversion of FI has the physical meaning of the variance (or covariance matrix in case of a multidimensional problem) \cite{frieden2004science}.

By definition, FI measures the overall sensitivity of the functional relationship $f(x|\mu)$ to changes in $\mu$, by weighting the sensitivity at each potential outcome $x$ with respect to the probability defined by $p_\mu(x) = f(x|\mu)$ \cite{schervish2012theory}. It can be expressed in the following form:
 \begin{equation}
 \mathfrak{F}(\mu)_{nm}=\int\limits_{M}^{} dx\left(\frac{\partial}{\partial \mu_n} \log f(x|\mu) \right)\left(\frac{\partial}{\partial \mu_m} \log f(x|\mu) \right)p_\mu(x).
\end{equation}

For a continuous set of 30 variables $\mu=\{\mu_A,\mu_W\}$, where $\mu_A=\{\bm{S}_{i}\}$ and $\mu_W=\{\bm{S}_{4+i}\}$ with $\bm{S}_l = \{I_l, Q_l, U_l, V_l\}$ for $i=\overline{1,4}$ (without $I_1$ and $I_5$). In general, the set of variables can be extended, for example, when working with a larger number of polarizations or when defining the parameters of auxiliary elements. For each particular measurement, we can assume that the intensity is normally distributed (with variance $\sigma^2$) and independent of other measurements. Additionally, we can assume that the variance is identical, regardless of the intensity value. Under these assumptions, the Fisher Information (FI) simplifies \cite{schervish2012theory}:
\begin{equation}\label{fisher}
    \mathfrak{F}_{nm}=\frac{1}{\sigma^2}\sum_{k,l,F,i}\frac{\partial}{\partial \mu_n}\left(     (A F(-\theta_i) M F (\theta_i)W)_{kl}\right)\frac{\partial}{\partial \mu_m}\left(    (A F(-\theta_i) M F (\theta_i)W)_{kl}\right).
\end{equation}
The diagonal elements of the inverse FI represent a lower bound on the variance of the corresponding variables, known as the Cramér–Rao bound (CRB) \cite{frieden2004science}:
\begin{equation}\label{cramer-rao}
    \Delta^2 \mu_n\ge(\mathfrak{F})^{-1}{}_{nn}.
\end{equation}
Thus, the existing of the inverse FI means principal possibility of restoring $\mu$. And moreover, the magnitude of diagonal elements reveals the relative error between the variables. This allows us to supplement the calibration procedure, as we can now study the calibration configuration theoretically, predict and deal with potential issues that may arise from it.

We can also determine how many measurements are needed for the unambiguous estimation of $A$ and $W$ for the chosen calibration configuration. This can be done in several ways for a given set of measurements: directly by checking the existence of the inverse FI, by calculating the eigenvalues of the FI (all of which must be non-zero), or by examining the matrix rank of the FI.

\subsection{Features of using common optical elements}

There are some comments regarding using a quarter-wave plate (QWP), a polarizer and a mirror as tools for calibration. In the case of using an ideal mirror as a reference sample, an ideal QWP and polarizer are insufficient for calibration. Indeed, the Mueller matrices of an ideal mirror ($M$), a QWP ($F_Q$) and a polarizer ($F_P$) are following:
\begin{eqnarray}
M= R(0,\pi/4,\pi),\\
F_Q=2\mathfrak{G}(0,0,1),\\
F_P=\mathfrak{G}(1,0,0),
\end{eqnarray}
and a measurement with an ideal QWP gives $M_{i_Q}=F_Q(-\theta_{i_Q})M F_Q(\theta_{I_Q})=J(2\theta_{i_Q})$ (with the Eq.~\eqref{MIS}-\eqref{FS}). This means physically that we rotate both the generator and the analyzer, and that can not provide an additional information about circular polarization (component $V_j$) compared to the measurement just with the mirror ($M_{i_M}=M$):
\begin{eqnarray}\label{IFQ1}
    &I_{i_Q}=I_k I_{4+l}+V_k V_{4+l}+(Q_k Q_{4+l}+U_k U_{4+l})\cos(4\theta_{i_Q})+(Q_{4+l} U_k-Q_k U_{4+l})\sin(4\theta_{i_Q}),\\
    &I_{i_M}=I_k I_{4+l}+V_k V_{4+l}+Q_k Q_{4+l}+U_k U_{4+l}.\label{IFQ2}
\end{eqnarray}
We can see in Eq.~\eqref{IFQ1}-\eqref{IFQ2} the same constant term $V_k V_{4+l}$, and there is an important feature in the calibration procedure of a setup with a BS. The use of an ideal QWP does not allow us to operate on the circular components of Stokes vector as freely as we can with the linear ones.

Nevertheless, there is always a small deviation from ideality in the Mueller matrices of a QWP and mirror in a real experiment, and it might seem that their presence is sufficient for carrying out the calibration. Unfortunately, the error in defining the components $V$ for these experiments will be significantly larger than for others. Specifically, this can be explicitly demonstrated by calculating the FI for nearly ideal cases: $\tilde{M}\equiv2\mathfrak{G}(0,\cos \beta,\sin \beta)$, with $\beta\approx\pi$ or $\tilde{F}_Q\equiv2\mathfrak{G}(0,\sin \gamma,\cos \gamma)$, with $\gamma\approx0$. The set of four measurements consists of one with an almost ideal mirror $\{M_1=\tilde{M}\}$, two with an ideal linear polarizer $\{M_{2,3}=F_P(-\theta)\tilde{M}F_P(\theta),\theta=0;\pi/4\}$ and the last with an ideal QWP $\{ M_4=F_Q(-\theta)\tilde{M} F_Q(\theta), \theta=\pi/4\}$. After inverting the FI, the variance of $V$-components grows as $\beta$ tends to $\pi$:
\begin{equation}\label{error_V48}
    \lim_{\beta\to\pi}\Delta^2 V\sim\sigma^2\frac{1}{(\pi-\beta)^2},
\end{equation}
while the variance for $I$-, $Q$- and $U$-components remains constant. A similar dependence appears, when we use an ideal mirror $M$ and nearly ideal QWP with the Muller matrix $\tilde{F}_Q$ (instead of $\tilde{M}$ and $F_Q$):
\begin{equation}\label{error_Q}
    \lim_{\gamma\to 0}\Delta^2 V\sim\sigma^2\frac{1}{\gamma^2}.
\end{equation}
If we measure a large number of angles for the polarizer and QWP $\theta_j=2\pi j/N $, $j=\overline{0,N-1}$, $N\gg 1$ (that means increasing the number of measurements to $N$) we can reduce variances of Eq.~\eqref{error_V48}-\eqref{error_Q} by approximately a factor of $N$. Taking into account the small value of $\sigma$, it is still possible to use non-ideal QWPs for calibration. However, despite the theoretical feasibility of this approach, we observe that the errors associated with different components of $A$ and $W$ vary by several degrees. This variation prevents us from considering this configuration as a standard calibration method. Therefore, we need to find a way to ensure that the errors for all components of the calibration matrices are of the same order of magnitude.

\subsection{Overcoming the difficulties in defining the $V$-components}

Because the use of polarizers between a sample and the BS allows for the definition of the linear components in the input and output Stokes vectors, but does not help with the circular ones, it seems logical to perform a measurement where the linear and circular components are swapped. We can achieve this separately for $W$ and $A$ by adding a QWP before ($F_{Q_1}=J(\Theta_1)F_Q J(-\Theta_1)$ at position A in Fig.~\ref{ris:schemes}(b)) and after ($F_{Q_2}=J(\Theta_2) F_Q J(-\Theta_2)$ at position B in Fig.~\ref{ris:schemes}(b)) the BS. Then, we need to extend the calibration variables $\zeta$ to include the ellipsometric angles $\{\psi_t,\Delta_t\,\psi_b,\Delta_b\}$ from the Mueller matrix $T=R(1,\Delta_t,\psi_t)$ and $B=R(1,\Delta_b,\psi_b)$, which correspond to the transmitted and reflected side of the BS respectively (transmission coefficients are not important for calibration problem in form Eq.~\eqref{MLM}, because they can be expressed as Eq.~\eqref{coef_MLM}). We then have the following sets of measurements:
\begin{itemize}    
    \item measurement without a polarizer and a QWP: 
    \begin{equation}\label{m1}
        \{M_{1}=T M B\},
    \end{equation}    
    \item measurements without a polarizer, but with a QWP $F_{Q_1}$ or $F_{Q_2}$: 
    \begin{equation}
       \{M_{2}(\Theta_1)=T M B F_{Q_1};  M_{3}(\Theta_2)=F_{Q_2} T M B\},
    \end{equation}
    \item measurements with a polarizer $F_P$, but without a QWP: 
    \begin{equation}
       \{M_{4}(\theta_{i_p})=T F_P(\theta_{i_p})M F_P(-\theta_{i_p}) B\}
    \end{equation} 
    \item measurements with a polarizer $F_P$ and with a QWP $F_{Q_1}$ or $F_{Q_2}$:
     \begin{equation}\label{m6}
      \{M_{5}(\theta_{i_{q_1}})=T F_P(\theta_{i_{q_1}})M F_P(-\theta_{i_{q_1}}) B F_{Q_1}; M_{6}(\theta_{i_{q_2}})= F_{Q_2} T F_P(\theta_{i_{q_2}})M F_P(-\theta_{i_{q_2}}) B\}.
    \end{equation} 
\end{itemize}
We will refer to this configuration as $Q-P-Q$ and the configuration with a polarizer and a QWP after the BS (in position C in Fig.~\ref{ris:schemes}(b)) as $-P/Q-$ for brevity.

\subsection{Monte Carlo simulation}

To demonstrate the effectivity of the $Q-P-Q$ calibration configuration, we compare its result from a numerical Monte Carlo simulation with the $-P/Q-$ configuration. We assume white Gaussian noise with a standard deviation $\sigma$ and define a signal to noise ratio:
\begin{equation}\label{SNR}
    SNR(\sigma)=10 \log_{10} \frac{1}{\sigma^2}.
\end{equation}
The accuracy of calibration procedure can be calculated with help of a root-mean-square error of estimator for $W=[w_{ij}]$ and $A=[a_{ij}]$:
\begin{equation}\label{RMSE}
    RMSE=\sqrt{\left\langle \sum_{ij}\left(a_{ij}-\mathfrak{a}_{ij}\right)^2\right\rangle+\left\langle \sum_{ij}\left(w_{ij}-\mathfrak{w}_{ij}\right)^2\right\rangle},
\end{equation}
where $W_0=[\mathfrak{w}_{ij}]$ and $A_0=[\mathfrak{a}_{ij}]$ are true values of calibration matrices without noise in tetrahedron form \cite{Goudail:09} (without coefficient $1/2$ to deal with normalized matrices and to avoid confusion in Eq.~\eqref{SNR}):
\begin{equation}\label{polar_tetrad}
W_0=A_0 {}^T=\left(
\begin{array}{cccc}
 1 & 1 & 1 & 1 \\
 \frac{1}{\sqrt{3}} & -\frac{1}{\sqrt{3}} & -\frac{1}{\sqrt{3}} & \frac{1}{\sqrt{3}} \\
 \frac{1}{\sqrt{3}} & -\frac{1}{\sqrt{3}} & \frac{1}{\sqrt{3}} & -\frac{1}{\sqrt{3}} \\
 \frac{1}{\sqrt{3}} & \frac{1}{\sqrt{3}} & -\frac{1}{\sqrt{3}} & -\frac{1}{\sqrt{3}} \\
\end{array}
\right).
\end{equation}
The polarizer is assumed to be modeled as $F_P=0.85\mathfrak{G}(0.999,0.01,0.01)$, the QWP as $F_Q=0.95*2\mathfrak{G}(0.01,0.01,0.999)$, the sample $M=R(1,\arctan(0.99),0.99 \pi)$ and the BS matrices $T=R(1,\arctan(0.99),0.01 \pi)$, $B=R(1,\arctan(0.99),0.99 \pi)$. The parameters of the optical elements do not affect the qualitative results of the simulation, but they were chosen to be close to those of mass-manufactured supplies.

\begin{figure}[h]
\begin{minipage}[b]{0.5\linewidth}
  \centering
  \includegraphics[width=1\columnwidth]{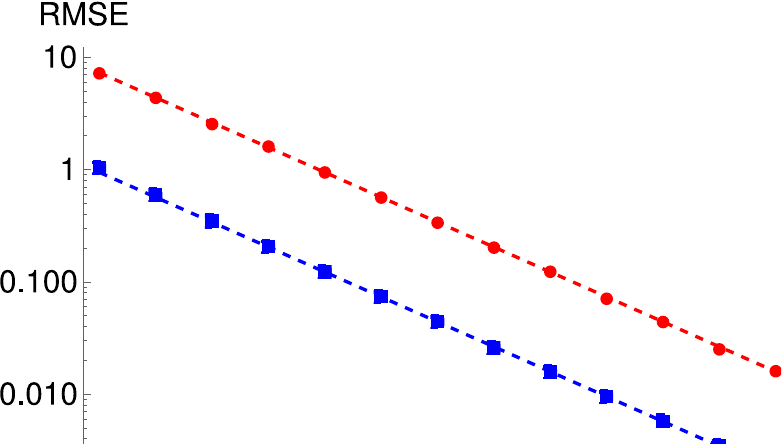}  \\
  (a)
\end{minipage}%
\begin{minipage}[b]{0.5\linewidth}
  \centering
   \includegraphics[width=1\columnwidth]{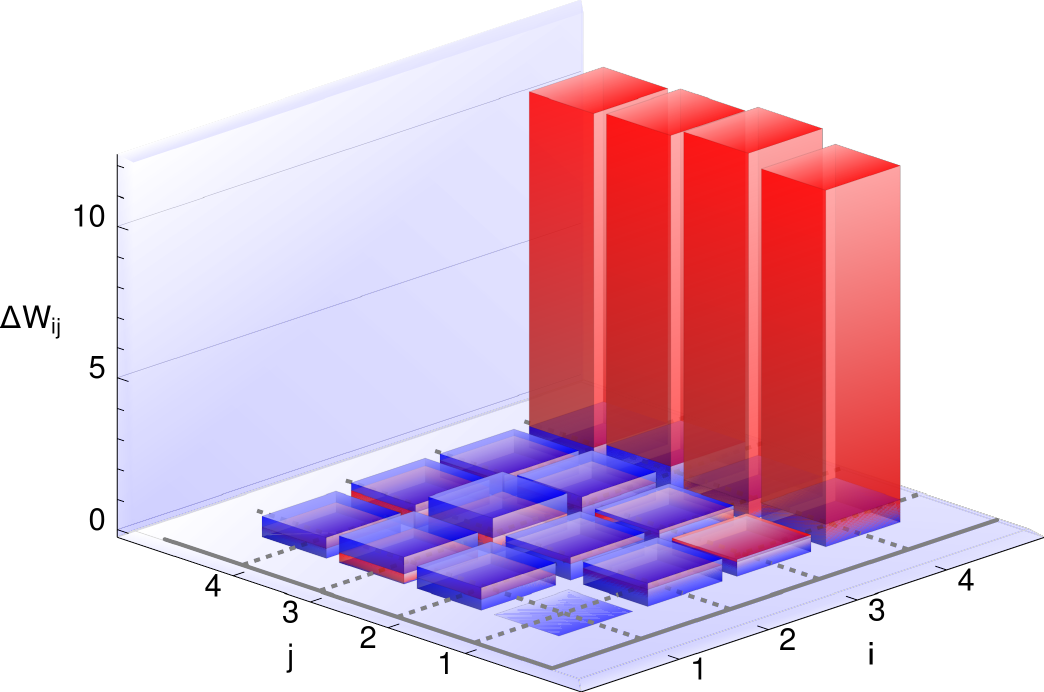} \\
  (b)
\end{minipage}
\caption{(a) Root mean square error function \eqref{RMSE} is in dependence of SNR for two calibration configurations: $Q-P-Q$ (blue points) and $-P/Q-$ (red points). Dashed lines show an analytical expression for minimal error from Cram{\'e}r–Rao bound. A number of simulation is 1000. (b) Standard deviation for $\Delta W_{ij}$ from Cram{\'e}r–Rao bound is shown for $Q-P-Q$ (blue barchart) and $-P/Q-$ (red barchart). }
\label{ris:simulation}
\end{figure}

We simulate 7 measurements for each calibration configuration (results are shown in Fig.~\ref{ris:simulation}(a)). For configuration $Q-P-Q$ (blue) we use measurements from Eq.  \eqref{m1}-\eqref{m6} with the following parameters: $\{M_1;M_2(\Theta_1=\pi/4);M_3(\Theta_2=\pi/4);M_4(\theta=0);M_5(\theta=\pi/4);M_6(\theta=0,\Theta_1=\pi/4);M_7(\theta=0,\Theta_2=\pi/4)\}$ (blue points), and the calculation of the FI of Eq.~\eqref{fisher} gives us $RMSE_{FI}\approx4.36\sigma$ (blue dashed line). For configuration $-P/Q-$ we use measurements: (i) without a polarizer or QWP $\{M_1=M\}$; (ii) three measurements with a polarizer $\{M_{2-4}=F_P(-\theta_{i_P})\tilde{M}F_P(\theta_{i_P}),\theta_{i_P}=0;\pi/9;\pi/4\}$; (iii)  three measurements with a QWP $\{ M_{5-7}=F_Q(-\theta_{i_Q})\tilde{M}F_Q(\theta_{i_Q}),\theta_{i_Q}=0;\pi/9;\pi/4\}$ (red points), with $RMSE_{FI}\approx33.65\sigma$ from the FI of Eq.~\eqref{fisher} (red dashed line). 

A standard deviation (as given by Eq.~\eqref{cramer-rao}) for all components of $W$ is shown in Fig.~\ref{ris:simulation}(b): blue for $Q-P-Q$ and red for $-P/Q-$. All of them are of the same order, except for the components of the circular polarizations in $-P/Q-$, as explained above. Also, a standard deviation for components of $A$ has the same features as $W^T$.

The FI demonstrates correct behavior in Fig.~\ref{ris:simulation}(a), and the difference between theoretical error from Eq.~\eqref{cramer-rao} and simulation results is a consequence of the general sense of the CRB compared to the particular case of restoring parameters by MLCM.

Thus, the described $Q$–$P$–$Q$ configuration provides a sufficient set of measurements for MLCM to produce errors of the same order for all components of the calibration parameters.

\section{A test for demonstrating the influence of a beamsplitter}

So, the calibration configuration $Q-P-Q$ allows us to recognize both the PSG and PSA calibration matrices, as well as the BS matrices: $T$ and $B$. Now, it is possible to examine the impact of the BS in a specific application.

We consider a simple sample consisting of a polarizer between the BS and the mirror. As a test we want to determine the orientation $\alpha$ of the polarizer. Without considering the impact of the BS, the calculated orientation $\alpha$ consists of the real orientation value $\theta$ and a systematic error $\delta$: $\alpha=\theta+\delta$. The log-likelihood function has the form similar to Eq.~\eqref{MLM}:
\begin{equation}\label{error}
    \mathfrak{l}(\alpha)=\frac{Tr\left[(A T^{(id)} P(-\alpha) M P (\alpha) B^{(id)} W)\cdot (A T P(-\theta) M P (\theta)B W)^T\right]^2}{Tr\left[(A T^{(id)}P(-\alpha) M P (\alpha)B^{(id)}W)\cdot (A T^{(id)} P(-\alpha) M P (\alpha)B^{(id)}W)^T\right]},
\end{equation}
with ideal BS elements  \cite{goldstein2017polarized} 
\begin{eqnarray}\label{MM_BS_id}
\begin{aligned}
    T^{(id)}=R(1,\pi/4,0),\\
   B^{(id)}=R(1,\pi/4,\pi),
\end{aligned}   
\end{eqnarray}
and with real ones $T=R(1,\arctan(P_t),\gamma_t)$ and $B=R(1,\arctan(P_b),\pi-\gamma_b)$, with $R$ given by Eq.~\eqref{mm_of_reflected}.

With following notations:
\begin{eqnarray}
    O=A T^{(id)}P(-\theta) M P (\theta)B^{(id)} W,\\
    X=\frac{\partial}{\partial{\theta}}O,\\
    K=A T P(-\theta) M P (\theta)B W-O,
\end{eqnarray}
we can find an analytic form of an approximate solution for $\delta$ from the equation $\partial \mathfrak{l}(\alpha)/\partial \alpha=0$, if we restrict ourselves to considering only the first term in the series for $\delta$ in Eq.~\eqref{error} due to its smallness. Additionally, the solution can be further simplified if we take into account that the Mueller matrices of the real BS are close to the ideal ones (we neglect terms of $R$ with powers greater than 2). The final solution is
\begin{eqnarray}\label{approx}
    \delta\approx\frac{Tr\left[X\cdot O^T\right]Tr\left[K\cdot O^T\right]-Tr\left[X\cdot K^T\right] Tr\left[O\cdot O^T\right] }{Tr\left[X\cdot O^T\right]^2- Tr\left[X\cdot X^T\right] Tr\left[O\cdot O^T\right]}.
\end{eqnarray}

For Stokes vectors in tetrahedron form Eq.~\eqref{polar_tetrad}, the error of Eq.~\eqref{approx} is equal to
\begin{equation}\label{difference}
    \delta^{(tet)}=\frac{1}{4}\left((1-P_t)+(1-P_b)\right)\sin 2\theta-\frac{1}{16}\left(\gamma_t^2+\gamma_b^2\right)\sin 4\theta.
\end{equation}
For the ideal case we have $P_t=P_b=1$ and $\gamma_t=\gamma_b=0$, and the error $\delta$ disappears. 

\section{Experimental validation}\label{sec:5}

We use a Mueller imaging polarimeter, illustrated in Fig.~\ref{ris:schemes}(b), to experimentally demonstrate the test described above. This device shares a common configuration with many polarimeters designed for backscattering geometry, consisting of two perpendicular arms. The first, horizontally oriented arm is designated for illumination and incorporates the PSG. The second, vertically oriented arm, used for detection, employs a conventional two-lens system and includes the PSA. The light source is a Picosecond Diode laser LDH-IB-730-B Taiko by PicoQuant, coupled into a single-mode fiber, operating at $\lambda=727.6$ nm. The laser beam generated by the source is collimated using a lens (LC) and directed onto the surface of the system under investigation via a nonpolarizing beamsplitter cube (Thorlabs BS017, 50:50). The PSG, which operates on the collimated beam, is capable of generating any arbitrary polarization state. This component includes a linear polarizer (Nanoparticle Linear Polarizer, Thorlabs, extinction ratio $1:100,000$) and two computer-controlled liquid crystal variable retarders (LCR, Meadowlark Optics). The fast axis of the first retarder is oriented at $\pi$/4 relative to the $x$- and $y$-axes of the laboratory frame, while the fast axis of the second retarder is aligned with the $y$-axis in the same frame. The polarized illumination beam is subsequently focused onto the sample surface using an objective lens ($L$) with a focal length of $f=60 mm$. 
The beam spot radius $1/e^2$ of $36 \mu m$ was determined by substituting the scattering system with a customized Pixelink camera for beam profiling and fitting a Gaussian profile to the recorded intensity distribution at various $z$ positions. The backscattered light traverses a two-lens system comprising the objective lens $L2$ and a second lens $L1$, identical to $L2$. An iris located at the back focal plane of $L2$ determines the numerical aperture (NA) of the imaging instrument. For this study, we employed an NA of $0.066$, which is sufficiently small to avoid imaging specular reflections when the illumination beam is slightly tilted ($~0.14 rad$). The PSA, situated between camera and two-lens system, consists of the same components as the PSG, but in reverse order. A CCD chip (ptGray Grasshopper, $16-bit$, $2448$ × $2048$ pixels) at the end of the detection arm is used to image the backscattered light with a magnification of approximately $1.2$. All measurements presented in this study were conducted at $\lambda = 727.6 nm.$

The experimental setup is quite robust, except for temperature and voltage fluctuations in the LCRs. These noise sources can be mitigated through appropriate stabilization measures. The camera operates at a low exposure time ($30$ $\mu$s), with no gain, zero black level, and $\gamma = 1$, to record intensity images in 16-bit format. The images are then converted to intensity values by summing over the camera region that contains the signal image. We also assume the default noise characteristics specified in the camera datasheet.

After carrying out the experiment, we can try to determine the orientation angle using two sets of calibration matrices defined via MLCM for $Q-P-Q$ configuration: $A_1 \equiv A \cdot T^{(id)}$, $W_1 \equiv B \cdot W^{(id)}$ and $A_2 \equiv A \cdot T$, $W_2 \equiv B \cdot W$ (for numerical details see Appendix~\ref{app:1}). Then, the expression in Eq.  \eqref{MLM} takes on a new meaning: it becomes relevant, but with only one unknown variable, the polarizer angle.

\begin{figure}[h]
\begin{minipage}[b]{0.5\linewidth}
  \centering
  \includegraphics[width=1\linewidth]{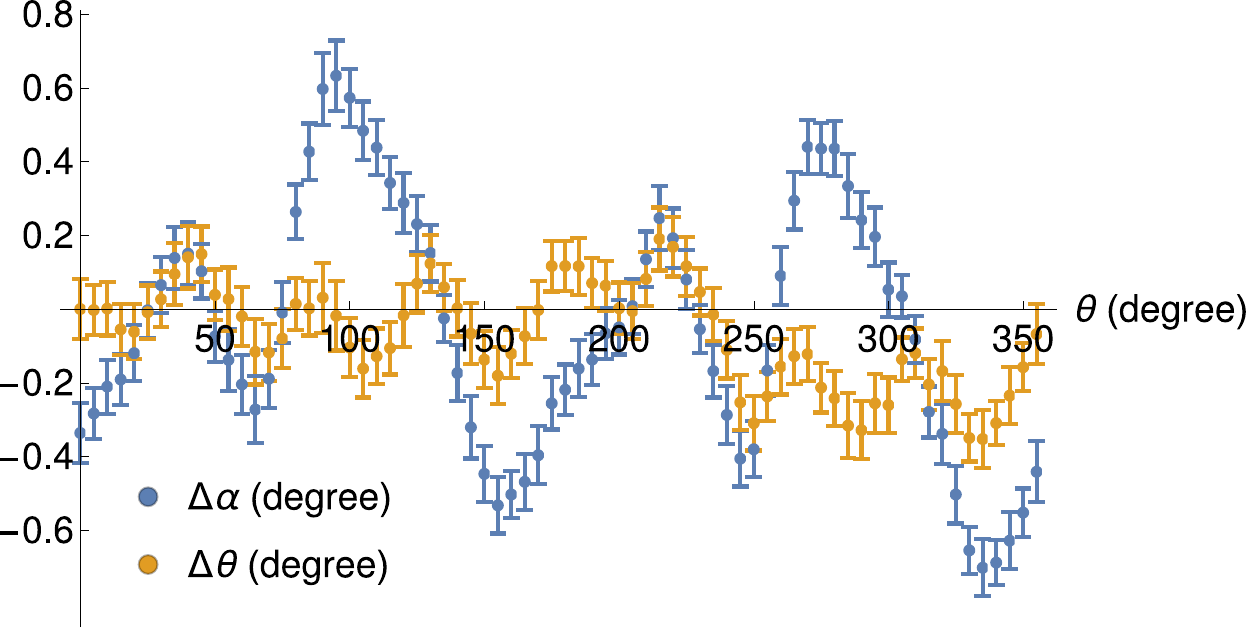}   
\end{minipage}%
\begin{minipage}[b]{0.5\linewidth}
  \centering
  \includegraphics[width=1\linewidth]{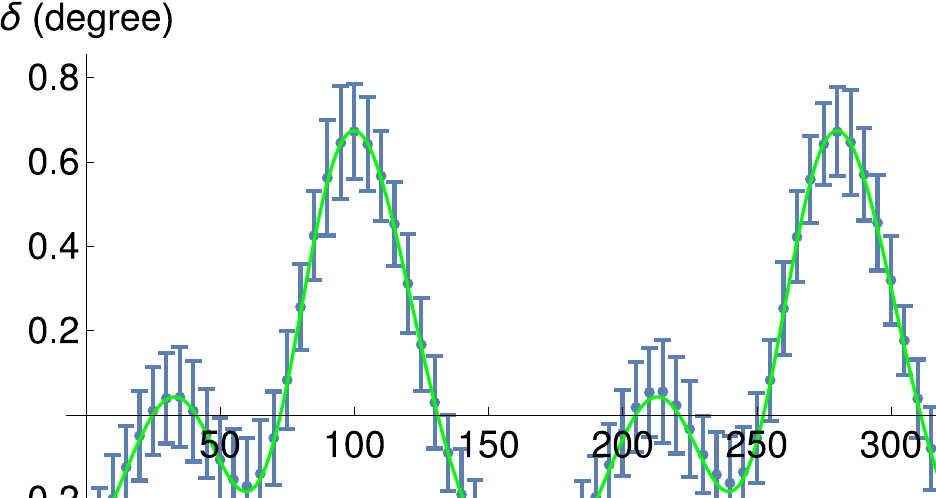} 
\end{minipage}
\newline
\begin{minipage}[b]{0.5\linewidth}
  \centering
  (a)
\end{minipage}%
\begin{minipage}[b]{0.5\linewidth}
  \centering
  (b)
\end{minipage}%
\caption{Results of the experiment for restoring the $\theta$ angle with 32 measurement for each polarizer position. Error bars show a standard deviation of the distributions. (a) shows the errors between the restored and known angles for two sets of calibration data: $\{A_1$; $W_1\}$ (without accounting for the impact of the BS) in blue, and $\{A_2$; $W_2\}$ (with the BS impact) in yellow. (b) shows the systematic error $\delta$ (blue line) and the fitting plot by the Eq.  \eqref{approx} (green line).}
\label{ris:plots}
\end{figure}

We measure 72 different polarizer positions with an equal step of 5 degrees.
The absolute difference ($\Delta\alpha=\alpha-\theta_0$ and $\Delta \theta=\theta-\theta_0$) between the reconstructed value of the orientation angle ($\alpha$ and $\theta$) and its known true value ($\theta_0$) is shown in Fig.~\ref{ris:plots}(a) for two sets of calibration matrices ($\{A_1$; $W_1\}$ and $\{A_2$; $W_2\}$) with a blue and yellow color correspondingly. In Fig.~\ref{ris:plots}(b) we see the $\delta=\alpha-\theta$ (blue points with error bar) as the difference between the blue and yellow plots in Fig.~\ref{ris:plots}(a), and the calculation of $\delta$ according to Eq.~\eqref{approx} is shown by a green line. Fig.~\ref{ris:plots}(b) demonstrates that the suggested approximation is good enough to estimate $\delta$. 

The systematic error $\delta$ arises purely from the use of unsuitable calibration matrices and can characterize the numerical impact of the BS. The magnitude of $\delta$ is on the order of half a degree, which is significant even for low-accuracy setups.

It is worth mentioning that, when the BS is neglected, any calibration method suitable for a polarimetric setup may be applied. However, this will obscure the pure impact of the BS, as the resulting systematic error will also include contributions from differences between calibration methods, for example, between extended ECM and MLCM, as discussed in \cite{hu2013maximum}.

{\section{Conclusion}}

In this paper, we demonstrate the importance of a correct calibration method for polarimetric backscattering setups based on a BS. Obviously, the optical elements of a BS (even those close to ideal) affect the accuracy of measurements. Despite the apparent insignificance of the BS’s influence, it cannot be neglected. To account for this, we describe a maximum likelihood calibration method adapted to a setup involving a BS. The so-called $Q-P-Q$ configuration yields errors of the same degree for all components of the calibration matrices, as shown numerically using the Fisher information framework. 

In general, the maximum likelihood calibration method can be applied to any setup architecture, provided the noise distribution is known, as this is required to define the likelihood function. The main drawback is the lack of explicit sensitivity to variations in input variables, especially in complex setups, and it can be addressed by concurrently checking of Fisher information. Even the simplifying assumption we made in this work regarding the Mueller matrices of the beamsplitter, where only two parameters are considered, can be easily relaxed by expanding the set of variables and solving the corresponding optimization problem. Altogether, these considerations confirm the maximum likelihood method as a universal tool for calibration purposes.

Additionally, we present a simple experimental test for restoring the polarizer orientation, which allows us to estimate the error introduced by ignoring the presence of a BS during the calibration of the setup.

\section*{Disclosures}
The authors declare no conflicts of interest.

\section*{Acknowledgment}
This work is supported by the Swiss National Science Foundation (Grant No. 200021\_212872).

\appendix
\section{Numerical values of the calibration matrices}\label{app:1}

Here, we describe the $A$, $W$, $T$, and $B$ matrices that we use to calculate  Eq.~\eqref{error} and Eq.~\eqref{approx}. We used two additional Stokes vectors to increase the accuracy of the setup. For this reason, the size of $A$ is 6 by 4, and the size of $W$ is 4 by 6:
\begin{equation*}
A=\left(
\begin{array}{cccc}
 0.5 & 0.472442 & -0.154061 & 0.0553524 \\
 0.492588 & -0.468833 & 0.101679 & -0.111804 \\
 0.500377 & 0.236969 & 0.440576 & 0.010761 \\
 0.492372 & -0.185684 & -0.455674 & -0.0177053 \\
 0.49444 & 0.0408987 & -0.00201927 & 0.492742 \\
 0.491526 & -0.0528187 & -0.107572 & -0.476693 \\
\end{array}
\right),
\end{equation*}
\begin{equation*}
W=\left(
\begin{array}{cccccc}
 0.5 & 0.470684 & 0.482135 & 0.485565 & 0.482711 & 0.487211 \\
 0.453892 & -0.44289 & 0.116669 & -0.146364 & -0.0174338 & 0.0967294 \\
 -0.206307 & 0.152238 & 0.467148 & -0.459702 & 0.0523683 & -0.145682 \\
 0.0376812 & 0.0470762 & -0.0248074 & -0.055002 & 0.479545 & -0.454747 \\
\end{array}
\right).
\end{equation*}

The parameters of the BS matrices $T=R(1,\arctan(P_t),\gamma_t)$ and $B=R(1,\arctan(P_b),\pi-\gamma_b)$ are calculated by the calibration procedure as $P_t=0.953517$, $P_b=1.02598$, $\gamma_t=-0.054225$ and $\gamma_b=-0.111118$. These are close to the values expected from the specifications of the manufacturer $P_t=0.957411$, $P_b=1.02101$. 
\bibliography{main.bib}

\end{document}